\documentclass[aps,preprint]{revtex4}
\usepackage{amssymb}

\usepackage{latexsym}
\usepackage{amsfonts}
\usepackage{graphicx}
\usepackage{epsfig}
\usepackage{amsmath}

\begin{document}

\title{A Generalized Preferential Attachment Model for
 Business Firms Growth Rates: II. Mathematical Treatment }

\author{S.~V.~Buldyrev$^1$, Fabio~Pammolli$^{2,3}$, 
Massimo~Riccaboni$^{2,3}$, Kazuko~Yamasaki$^4$, Dongfeng~Fu$^5$, 
Kaushik~Matia$^5$, H.~E.~Stanley$^5$ }


\affiliation{ $^1$Department of Physics,~Yeshiva University, 500 West 185th Street,~New York, NY 10033 USA \\ 
$^{2}$Faculty of Economics, University of Florence, Milan, Italy \\
$^{3}$IMT Institute for Advanced Studies, Via S.~Micheletto 3, Lucca, 55100 Italy \\ 
$^4$Tokyo University of Information Sciences, Chiba City 265-8501 Japan$^1$ \\
$^5$Center for Polymer Studies and Department of Physics, Boston University, Boston, MA 02215 USA }

\begin{abstract}
  We present a preferential attachment growth model to obtain the
  distribution $P(K)$ of number of units $K$ in the classes which may
  represent business firms or other socio-economic entities.  We found
  that $P(K)$ is described in its central part by a power law with an
  exponent $\varphi=2+b/(1-b)$ which depends on the probability of entry of new
  classes, $b$. In a particular problem of city
  population this distribution is equivalent to the well known Zipf
  law. In the absence of the new classes entry, the
  distribution $P(K)$ is exponential.   
  Using analytical form of $P(K)$ and assuming proportional growth
  for units, we derive $P(g)$, the distribution of business firm
  growth rates. The model predicts that $P(g)$ has a Laplacian cusp
  in the central part and asymptotic power-law tails with an exponent
  $\zeta=3$. We test the analytical expressions derived using
  heuristic arguments by simulations. The model might also explain the
  size-variance relationship of the firm growth rates.
\end{abstract}


\maketitle

\section{Introduction}

Gibrat~\cite{Gibrat30, Gibrat31}, building upon the work of the astronomer
Kapteyn~\cite{Kapteyn16}, assumed the expected value of the growth rate of a
business firm's size to be proportional to the current size of the firm,
which is called ``Law of Proportionate Effect''~\cite{Zipf49,Gabaix99}. Several
models of proportional growth have been subsequently introduced in economics
in order to explain the growth of business firms~\cite{Steindl65, Sutton97,
Kalecki45}. Simon and co-authors~\cite{Simon55, Simon58, Simon75, Simon77}
extended Gibrat's model by introducing an entry process according to which
the number of firms rise over time. In Simon's framework, the market consists of a
sequence of many independent ``opportunities'' which arise over time, each of
size unity. Models in this tradition have been challenged by many
researchers~\cite{Stanley96, Lee98, Stanley99, Bottazzi01, Matia04} who found
that the firm growth distribution is not Gaussian but displays a tent shape.

Here we introduce a mathematical framework that provides an unifying explanation
for the growth of business firms based on the number and size distribution of
their elementary constituent
components~\cite{Amaral97,Sergey_II,Sutton02,DeFabritiis03,Amaral98,Takayasu98,Canning98,
Buldyrev03}. Specifically we present a model of proportional growth in both
the number of units and their size and we draw some general implications on
the mechanisms which sustain business firm growth~\cite{Simon75, Sutton97,
Kalecki,Mansfield,Hall,DeFabritiis03}.  According to the model, the
probability density function (PDF) of growth rates, $P(g)$ is Laplace \cite{Kotz01} in the
center~\cite{Stanley96} with power law tails~\cite{Reed01,Reed02} decaying as
$g^{-\zeta }$ where $\zeta=3$.

Two key sets of assumptions in the model are described in subsections A (the number of
units $K$ in a class grows in proportion to the existing number of units) and B
(the size of each unit fluctuates in proportion to its size). Our goal is to
first find $P(K)$, the probability distribution of the number of units in the
classes at large $t$, and then find $P(g) using the convolution of $P(K) and the conditional
distribution of the class growth rates $P(g|K)$, which for large $K$  converges to a Gaussian..

\section{Analytical Results}

\subsection{The Proportional Growth of Number of Units}~\label{section_PK}
The first set of assumptions \cite{Kazuko} is: 
\begin{itemize}
\item[{(A1)}] Each class $\alpha$ consists of $K_{\alpha}(t)$ number of
  units. At time $t=0$, there are
  $N(0)$ classes consisting of $n(0)$ total number of units. The initial
  average number of units in a class is thus $n(0)/N(0)$.

\item[{(A2)}] At each time step a new unit is created. Thus the number
  of units at time $t$ is $n(t)=n(0)+t$.

\item[{(A3)}] With birth probability $b$, this new unit is assigned to a new
  class, so that the average number of classes at time $t$ is $N(t)=N(0)+bt$.

\item[{(A4)}] With probability $1-b$, a new unit is assigned to an existing
  class $\alpha$ with probability $P_{\alpha}=(1-b)K_{\alpha}(t)/n(t)$, so
  $K_{\alpha}(t+1) = K_{\alpha}(t)+1$.
\end{itemize}

This model can be generalized to the case when the units are born at
any unit of time $t'$ with probability $\mu$, die with probability
$\lambda$, and in addition a new class consisting of one unit can be
created with probability $b'$ \cite{Kazuko}.  This model can be
reduced to the present model if one introduce time
$t=t'(\mu-\lambda+b')$ and probability $b=b'/(\mu-\lambda+b')$.

Our goal is to find $P(K)$, the probability distribution of the number of
units in the classes at large $t$. This model in two limiting cases (i)
$b=0$, $K_{\alpha}=1$ $(\alpha =1,2 \ldots N(0))$ and (ii) $b\neq 0$,
$N(0)=1$, $n(0)=1$ has exact analytical solutions
$P(K)=N(0)/t(t/(t+N(0)))^K (1+O(1/t))$~\cite{Johnson,Kotz2000} and
$\lim\limits_{t\to\infty}P(K)=(1+b)\Gamma(K)
\Gamma(2+b)/\Gamma(K+2+b)$~\cite{Reed04} respectively, In general,
an exact analytical solution of this problem cannot be presented in a
simple close form. Accordingly, we seek for an approximate mean-field
type~\cite{book} solution which can be expressed in simple integrals
and even in elementary functions in some limiting cases. 
First we will present a known solution of the preferential attachment model
in the absence of the influx of new classes~\cite{Cox}:
\begin{equation}
P_{\rm old}(K)=\lambda^K\frac{1}{K(t)-1}\approx \frac{1}{K(t)}\exp(-K/K(t))[1+O(t^{-1})],
\label{P_K_old}
\end{equation}
where $\lambda=1-1/K(t)$ and $K(t)=[n(0)+t]/N(0)$ is the average number of units in the old
classes at time $t$. Note that the form of the distribution of units in the
old classes remains unchanged even in the presence of the new classes, whose
creation does not change the preferential attachment mechanism of the old
classes and affects only the functional form of $K(t)$.

Now we will treat the problem in the presence of the influx of the new classes.
Assume that
at the beginning there are $N(0)$ classes with $n(0)$ units. Because
at every time step, one unit is added to the system and a new class is
added with probability $b$, at moment $t$ there are 
\begin{equation}
n(t)= n(0) + t
\end{equation}
units and approximately
\begin{equation}
N(t)=N(0)+bt
\end{equation}
classes, among which there are approximately $bt$ new
classes with $n_{new}$ units and $N(0)$ old classes with $n_{old}$
units, such that 
\begin{equation}
n_{old} + n_{new} = n(0) + t.
\end{equation}
Because of the preferential attachment assumption~(A4), we can
write, neglecting fluctuations~\cite{book} and assuming that $t$, $n_{old}$, and $n_{new}$ are continuous
variables:
\begin{eqnarray}
\frac{dn_{new}}{dt} & = & b + (1-b) \frac{n_{new}}{n(0) + t}, \\
\frac{dn_{old}}{dt} & = & (1-b) \frac{n_{old}}{n(0)+t}.
\end{eqnarray}
Solving the second differential equation and taking into account initial
condition $n_{old}(0)=n(0)$, we obtain $n_{old}(t) = (n(0) + t)^{1-b} \,\,
n(0)^b$. Analogously, the number of units at time $t$ in the classes existing
at time $t_0$ is 
\begin{equation}
n_{e}(t_0,t) = (n(0) + t)^{1-b}(n(0) + t_0)^b
\label{n_e}
\end{equation}
where the subscript `e' means ``existing''. Accordingly, the average number of units in old classes is
\begin{eqnarray}
K(t) = \frac{ n_{old}(t) }{N(0)} = \frac{ (n(0) + t)^{1-b} }{N(0)}\,\, n(0)^b.
\label{mean_K_old}
\end{eqnarray}
Thus according to Eq. (\ref{P_K_old}), the distribution of units in the 
old classes is
\begin{eqnarray}
P_{old}(K) \approx \frac{N(0)}{(n(0) + t)^{1-b} n(0)^b}\,\,\,\exp\left( -
\frac{K\,N(0)}{(n(0) + t)^{1-b}n(0)^b}\right). \label{p_old_class}
\end{eqnarray}
and the contribution of the old classes to the distribution
of all classes is 
\begin{equation}
\tilde{P}_{old}(K)= P_{old}(K) N(0)/(N(0)+bt).
\label{tildeP_K_old}
\end{equation}
The number of units in the classes that appear at $t_0$ is $b\,dt$ and the
number of these classes is $b\,dt$. Because the probability that a class
captures a new unit is proportional to the number of units it has already
gotten at time $t$, the number of units in the classes that appear at time
$t_0$ is 
\begin{equation}
n_{new}(t_0,t) = n_{e}(t_0,t) bdt/[n(0) + t_0]. 
\end{equation}
The average number
of units in these classes is 
\begin{equation}
K(t_0,t)=n_{new}(t_0,t)/b\,dt=(n(0) +
t)^{1-b}/(n(0)+t_0)^{1-b}. 
\end{equation}
Assuming that the distribution of units in these
classes is given by a continuous approximation (\ref{P_K_old}) we have
\begin{eqnarray}
P_{new}(K, t_0) \approx \frac{1}{K(t_0,t)}\,\exp\left(-K/K(t_0,t) \right).
\end{eqnarray}
Thus, their contribution to the total distribution is
\begin{eqnarray*}
\frac{b\,dt_0}{N(0) + b\,t}\,\frac{1}{K(t_0,t)}\,\exp\left(-K/K(t_0,t) \right)
\end{eqnarray*}
The contribution of all new classes to the distribution $P(K)$ is
\begin{eqnarray}
\tilde{P}_{new}(K) \approx \frac{b}{N(0) + b\,t}\int_0^t
\frac{1}{K(t_0,t)}\,\exp\left(-K/K(t_0,t) \right)\, dt_0.
\label{tildeP_K_new}
\end{eqnarray}

If we let $y = K/K(t_0,t)$ then $\tilde{P}_{new}(K)=P_{new}(K)bt/(N(0)+bt)$ where 
\begin{eqnarray}  
P_{new}(K) & \approx & \frac{n(0)/t+1}{1-b}\,\, K^{\left(-\frac{1}{1-b} -
1\right)}\,\, \int^K_{K'} e^{-y} \,\,y^{\frac{1}{1-b}} \,\,dy.
\label{tilde_p_new_K}
\end{eqnarray}
and the low limit of integration, $K'$
is given by
\begin{equation}
K'=K\left(\frac{n(0)}{n(0)+t}\right)^{1-b}
\end{equation}
Finally the distribution of units in all classes is given by
 \begin{equation}
 P(K)= \frac{N(0)}{N(0) + bt}\,P_{old}(K)\,+\,\frac{bt}{N(0)+bt}\,P_{new}(K).
 \label{p_K_final}
 \end{equation}
Now we investigate the asymptotic behavior of the distribution in Eq.~(\ref{tilde_p_new_K})
and show that it can be described by the Pareto power law tail with an exponential
cut-off.

1. At fixed $K$ when $t \rightarrow \infty$, we have $K'\to 0$, thus
\begin{eqnarray}
P_{new}(K) & = & \frac{1}{1-b}\,\, K^{-\frac{1}{1-b} - 1} \int_0^K e^{-y}\,\,
y^{\frac{1}{1-b}} \,\, dy, \notag \\ & = & \frac{1}{1-b}\,
\left[\Gamma\left(1+\frac{1}{1-b}\right) - \int_K^\infty
e^{-y}\,\,y^{\frac{1}{1-b}}\,\, dy \right]\,\,K^{-1-\frac{1}{1-b}}.
\end{eqnarray}
As $K \rightarrow \infty$, $P_{new}(K)$ converges to a finite value:
\begin{eqnarray}
P_{new}(K) & = & K^{-1-\frac{1}{b}} \left( \frac{1}{1-b}\right)
\,\,\Gamma\left(1+\frac{1}{1-b}\right).
\end{eqnarray}
Thus for large $K\gg 1$, but such that $K'\ll 1$ or $K\ll (1+t/n(0)^{1-b}$, we have an
approximate power-law behavior:
\begin{equation}
P_{new}(K)\sim K^{-\varphi},
\end{equation}
where $\varphi=2+b/(1-b)\ge 2$.  

As $K\rightarrow 0$,
\begin{eqnarray}
P_{new}(K) & = & \frac{1}{1-b}\,\, K^{\left(-\frac{1}{1-b} - 1\right)}\,\,%
\frac{K^{\left(1+\frac{1}{1-b}\right)}}{1+\frac{1}{1-b}}  = \frac{1}{2-b}.
\end{eqnarray}
2. At fixed $t$ when $K \rightarrow \infty$, we use the partial integration to
   evaluate the incomplete $\Gamma$ function:
\begin{eqnarray*}
\,\,\,\,\,\, \int_x^{\infty}e^{-y}\,\,y^{\alpha}\,\,dy & = &
-e^{-y}\,\,y^\alpha |_x^\infty + \alpha\int_x^\infty
e^{-y}\,\,y^{\alpha-1}\,\,dy \approx e^{-x}\,\,x^{\alpha}.
\end{eqnarray*}
Therefore, from Eq.~(\ref{tilde_p_new_K}) we obtain
\begin{eqnarray}
\tilde{P}_{new}(K) &\approx & \frac{n(0)+t}{N(0)+bt}\,\,\frac{b}{1-b} \,\,K^{-\frac{1}{1-b}
  -1} \int^\infty_{K\left(\frac{n(0)}{n(0)+t} \right)^{1-b}} \,\,e^{-y}\,\,y^{
  \frac{1}{1-b}}\,\,dy, \notag \\ &=&
\frac{n(0)}{N(0)+bt}\,\,\frac{b}{1-b}\,\,\frac{1}{K}\,\,\exp\left(-K\left(
\frac{ n(0)}{n(0)+t}\right)^{1-b}\right),
\end{eqnarray}
which always decays faster than Eq.~(\ref{p_old_class}) because $n(0)\geq N(0)$ and there is an additional factor
$K^{-1}$ in front of the exponential. Thus the behavior of the
distribution of all classes is dominated for large $K$ by the exponential decay of the distribution
of units in the old classes.

\begin{figure}
\centering
\includegraphics[scale=0.25,angle=-90]{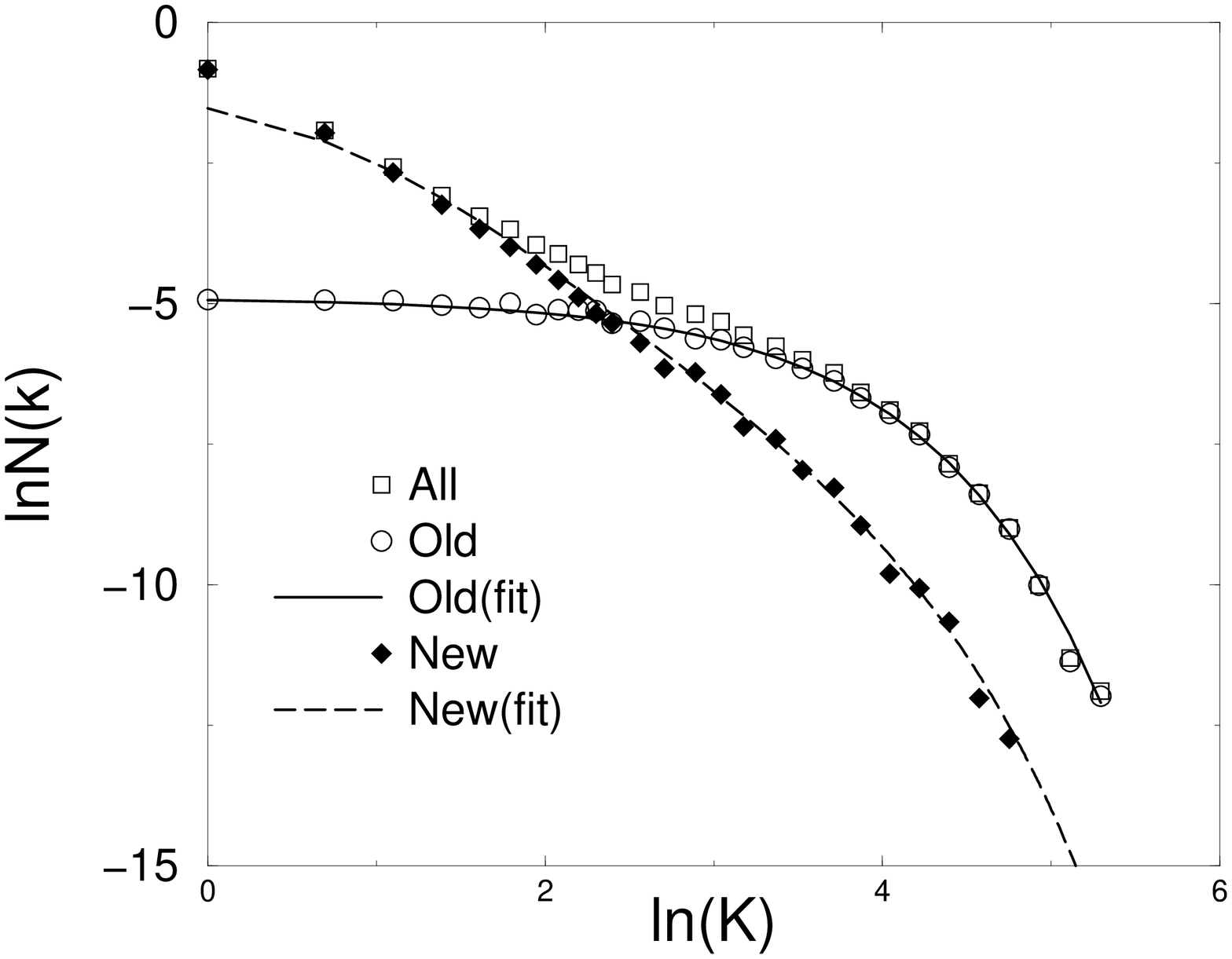}
\caption{Comparison of the distributions $P(K)$ for the new and old classes obtained by numerical simulations of the model with the predictions
of Eq. (\ref{tildeP_K_new}) and Eq. (\ref{tildeP_K_old}) respectively. 
For large $K$ the agreement is excellent.
The discrepancy exists only for $\tilde{P}_{new}$
at small $K$, e.g. Eq. (\ref{tildeP_K_new}) significantly underestimates the $\tilde{P}_{new}(1)$ and 
$\tilde{P}_{new}(2)$.}
\label{pk-new}
\end{figure}

Note that Eq.~(\ref{p_old_class}) and Eq.~(\ref{tilde_p_new_K}) are not exact
solutions but continuous approximations which assume $K$ is a real
number. This approximation produces the most serious discrepancy for small $K$.
To test this approximation, we perform numerical simulations of the model for $b=0.1$,
$N(0)=n(0)=10000$ and $t=400000$. The results are presented in Fig.\ref{pk-new}.
While the agreement is excellent for large $K$, Eq. (\ref{tilde_p_new_K}) significantly
underestimates the value of $\tilde{P}_{new}(K)$ for $K=1$ and $K=2$. Note that in
reality the power-law behavior of $\tilde{P}_{new}(K)$ extends into the region of very
small $K$. 

\subsection{The Proportional Growth of Size of Units}

The second set of assumptions of the model is:

\begin{itemize}
\item[{(B1)}] At time $t$, each class $\alpha$ has $K_{\alpha}(t)$ units of
size $\xi_i(t)$, ${i=1,2,...K_{\alpha}(t)}$ where $K_{\alpha}$ and $\xi_i >
0$ are independent random variables taken from the distributions
$P(K_{\alpha})$ and $P_{\xi}(\xi_i)$ respectively. $P(K_{\alpha})$ is defined
by Eq.~(\protect\ref{p_K_final}) and $P_{\xi}(\xi_i)$ is a given distribution
with finite mean and standard deviation and $\ln\xi_i$ has finite mean
$\mu_{\xi}=\langle\ln\xi_i\rangle$ and variance
$V_{\xi}=\langle(\ln\xi_i)^2\rangle-\mu_{\xi}^2$. The size of a class is
defined as $S_{\alpha}(t)\equiv \sum_{i=1}^{K_{\alpha}}\xi_i(t)$.
\item[{(B2)}] At time $t+1$, the size of each unit is decreased or
  increased by a random factor $\eta_i(t)>0$ so that
\begin{equation}
\xi_i(t+1)=\xi_i(t)\,\eta_i(t),
\end{equation}
where $\eta_i(t)$, the growth rate of unit $i$, is independent random
variable taken from a distribution $P_\eta(\eta_i)$, which has finite mean and standard deviation.
We also assume that $\ln \eta_i$ has finite mean
$\mu_{\eta}\equiv\langle\ln\eta_i\rangle$ and variance
$V_{\eta}\equiv\langle(\ln\eta_i)^2\rangle-\mu_{\eta}^2$.
\end{itemize}

Let us assume that due to the Gibrat process, both the size and growth of units ($\xi_i$ and $\eta_i$
respectively) are distributed lognormally
\begin{equation}
p(\xi_i)={\frac{1}{\sqrt{2\pi V_\xi}}}\,\,{\frac{1}{\xi_i}}
\,\,\exp\left(-(\ln\xi_i-m_\xi)^2/2V_\xi\right),
\end{equation}
\begin{equation}
p(\eta_i)={\frac{1}{\sqrt{2\pi V_\eta}}}\,\,{\frac{1}{\eta_i}}
\,\,\exp\left(-(\ln\eta_i-m_\eta)^2/2V_\eta\right).
\end{equation}
If units grow according to a multiplicative process, the size of units $
\xi_i^{\prime}=\xi_i\eta_i$ is distributed lognormally with $
V_{\xi^{\prime}}=V_\xi+V_\eta$ and $m_{\xi^{\prime}}=m_\xi + m_\eta$.

The $n^{\mbox{\scriptsize th}}$ moment of the variable $x$
distributed lognormally is given by
\begin{eqnarray}
\mu_x(n) &=& \int_0^\infty{\frac{1}{\sqrt{2\pi
V}}}\,{\frac{x^n}{x}}\,dx\,\exp\left(-(\ln x-m)^2/2V\right) \,\,=\,\,
\exp\left(nm_x+n^2V_x/2\right).
\label{mun}
\end{eqnarray}
Thus, its mean is $\mu_x \equiv \mu_x(1)= \exp(m_x+V_x/2)$ and its variance is
$\sigma_x^2\equiv \mu_x(2)-\mu_x(1)^2=\mu_x(1)^2\,(\exp(V_x)-1)$.

Let us now find the distribution of $g$ growth rate of classes. It is defined
as
\begin{equation}
g\equiv \ln{\frac{S(t+1)}{S(t)}}=\ln\sum_{i=1}^K\xi_i^{\prime}-\ln\sum_{i=1}^K\xi_i.
\end{equation}
Here we neglect the influx of new units, so
$K_{\alpha}=K_{\alpha}(t+1)=K_{\alpha}(t)$. 

The resulting distribution of thegrowth rates of all classes is determined by
 \begin{equation}
 P(g) \equiv \sum_{K=1}^{\infty}P(K)P(g|K),
 \label{P_g_sum}
 \end{equation}
 where $P(K)$ is the distribution of the number of units in the classes,
 computed in the previous stage of the model and $P(g|K)$ is the conditional
 distribution of growth rates of classes with given number of units determined
 by the distribution $P_{\xi}(\xi)$ and $P_{\eta}(\eta)$. 

Now our goal is to find an analytical approximation for $P(g|K)$.
 According to the central limit theorem, the sum of $K$ independent random
 variables with mean $\mu_{\xi}\equiv \mu_{\xi}(1)$ and finite variance
 $\sigma_{\xi}^2$ is
\begin{equation}
\sum_{i=1}^K\xi_i=K\mu_\xi+\sqrt K\nu_K,
\label{nu}
\end{equation}
where $\nu_K$ is the random variable with the distribution converging to
Gaussian
\begin{equation}
\lim_{K\to\infty}P(\nu_K)\to{\frac{1}{\sqrt{2\pi\sigma_\xi^2}}}\,\,
\exp\left(-\nu_K^2/2\sigma_\xi^2\right).
\end{equation}
Accordingly, we can replace $\ln(\sum_{i=1}^K\xi_i)$ by its Tailor's expansion 
$\ln K + \ln \mu_\xi+\nu_K/(\mu_\xi\sqrt{K})$, neglecting the terms of order $K^{-1}$.
Because $\ln \mu_{\eta}=m_\eta+V_\eta/2$ and $\ln
\mu_{\xi^{\prime}}=\ln \mu_{\xi} + \ln \mu_{\eta}$ we have

\begin{eqnarray}
g\equiv \ln S(t+1)-\ln S(t) &=& \ln(K\mu_{\xi^{\prime}})+{\frac{
\nu_K^{\prime}}{\sqrt K\mu_{\xi^{\prime}}}}- \ln(K\mu_\xi) - {\frac{\nu_K}{
\sqrt K\mu_{\xi}}}, \notag \\ &=&
m_\eta+{\frac{V_{\eta}}{2}}+{\frac{\nu_K^{\prime}\mu_\xi-\nu_K\mu_{\xi^{\prime}}}{\sqrt
K\mu_\xi\mu_{\xi^{\prime}}}}.
\label{g_m_V}
\end{eqnarray}
For large $K$ the last term in Eq.~(\ref{g_m_V}) is the difference of two
Gaussian variables and that is a Gaussian variable itself. Thus for large $K$, $g$ converges
to a Gaussian with mean, $m=m_\eta+V_\eta/2$, and certain standard deviation which we must find.

In order to do this, we rewrite
\begin{equation*}
\frac{\nu_K^{\prime}}{\sqrt K\,\mu_{\xi^{\prime}}} = \frac{
\sum_{i=1}^K(\xi_i^{\prime}-\mu_{\xi^{\prime}})}{K\,\mu_{\xi^{\prime}}},
\end{equation*}
and
\begin{equation*}
\frac{\nu_K}{\sqrt K\,\mu_\xi}=\frac{\sum_{i=1}^K(\xi_i - \mu_\xi)}{
K\,\mu_\xi}.
\end{equation*}
Thus
\begin{eqnarray}
g &=&
m_\eta+{\frac{V_\eta}{2}}\,+\,{\frac{\sum_{i=1}^K\xi_i(\eta_i\mu_\xi-\mu_{\xi^{\prime}})}{K\mu_\xi\mu_{\xi^{\prime}}}},
\notag \\ &=& m_\eta+{\frac{V_\eta}{2}}+
{\frac{\sum_{i=1}^K\xi_i(\eta_i-\mu_\eta)}{K\mu_{\xi^{\prime}}}}.  
\label{g}
\end{eqnarray}

Since $\mu_{\xi^{\prime}} = \mu_{\xi} \mu_{\eta}$, the average of each term
in the sum is $\mu_{\xi^{\prime}}-\mu_\xi\,\mu_\eta=0$. The variance of each
term in the sum is $\langle(\xi_i\,\eta_i)^2\rangle-\langle
2\xi_i^2\,\eta_i\,\mu_\eta\rangle +\langle\xi_i^2\,\mu_{\eta}^2\rangle$ where
$\xi_i\eta_i$, $\xi_i^2\eta_i$ and $\xi_i^2$ are all lognormal independent
random variables. Particularly, $(\xi_i\eta_i)^2$ is lognormal with $V =
4V_\eta+4V_\xi$ and $m=2m_\eta+2m_\xi$; $\xi_i^2\eta_i$ is lognormal with
$V = 4V_\xi+V_\eta$ and $m = 2m_\xi+m_\eta$; $\xi_i^2$ is lognormal with $V =
4V_\xi$ and $m = 2m_\xi$. Using Eq.~(\protect\ref{mun})
\begin{subequations}
\begin{align}
\langle(\xi_i\eta_i)^2\rangle &=
\exp(2m_\eta+2m_\xi+2V_\eta+2V_\xi),\label{mean_all_a} \\
\langle\xi_i^2\eta_i\rangle &=
\exp(m_\eta+2m_\xi+2V_\xi+V_\eta/2),\label{mean_all_b} \\
\langle\xi_i^2\rangle &= \exp(2m_\xi+2V_\xi).
\label{mean_all_c}
\end{align}
\end{subequations}
Collecting all terms in Eqs.~(\ref{mean_all_a}-\ref{mean_all_c}) together and
using Eq.~(\ref{g}) we can find the variance of $g$:
\begin{eqnarray}
\sigma^2 &=&
\frac{K\,\exp(2m_\xi+2V_\xi+2m_\eta+V_\eta)(\exp(V_\eta)-1)}{K^2\exp(2m_\xi+V_\xi+2m_\eta+V_\eta)},
\notag \\ &=& \frac{1}{K}\exp(V_\xi)\,(\exp(V_\eta)-1).
\end{eqnarray}

Therefore, for large $K$, $g$ has a Gaussian distribution
\begin{equation}
P(g|K)={\frac{\sqrt K}{\sqrt{2\pi V}}}\,\exp\left(-\frac{(g-m)^2K}{2V}\right),
\label{P_g_large_K}
\end{equation}
where 
\begin{equation}
m = m_\eta+ V_\eta/2
\label{e.m}
\end{equation}
 and 
\begin{equation}
V\equiv K\sigma^2 = \exp(V_\xi)(\exp(V_\eta)-1).
\label{e.V}
\end{equation}

\begin{figure}
\centering
\includegraphics[scale=0.25,angle=-90]{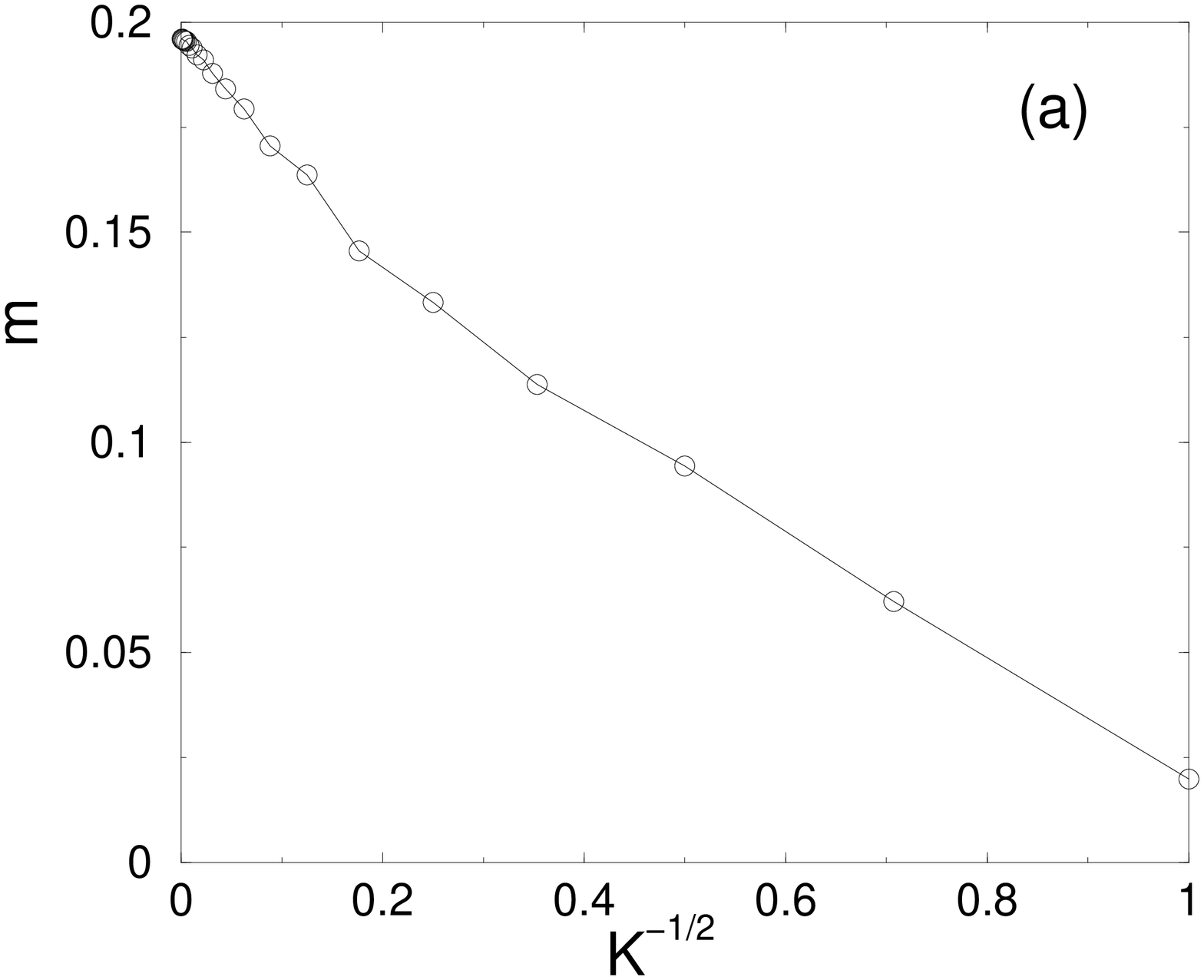}
\includegraphics[scale=0.25,angle=-90]{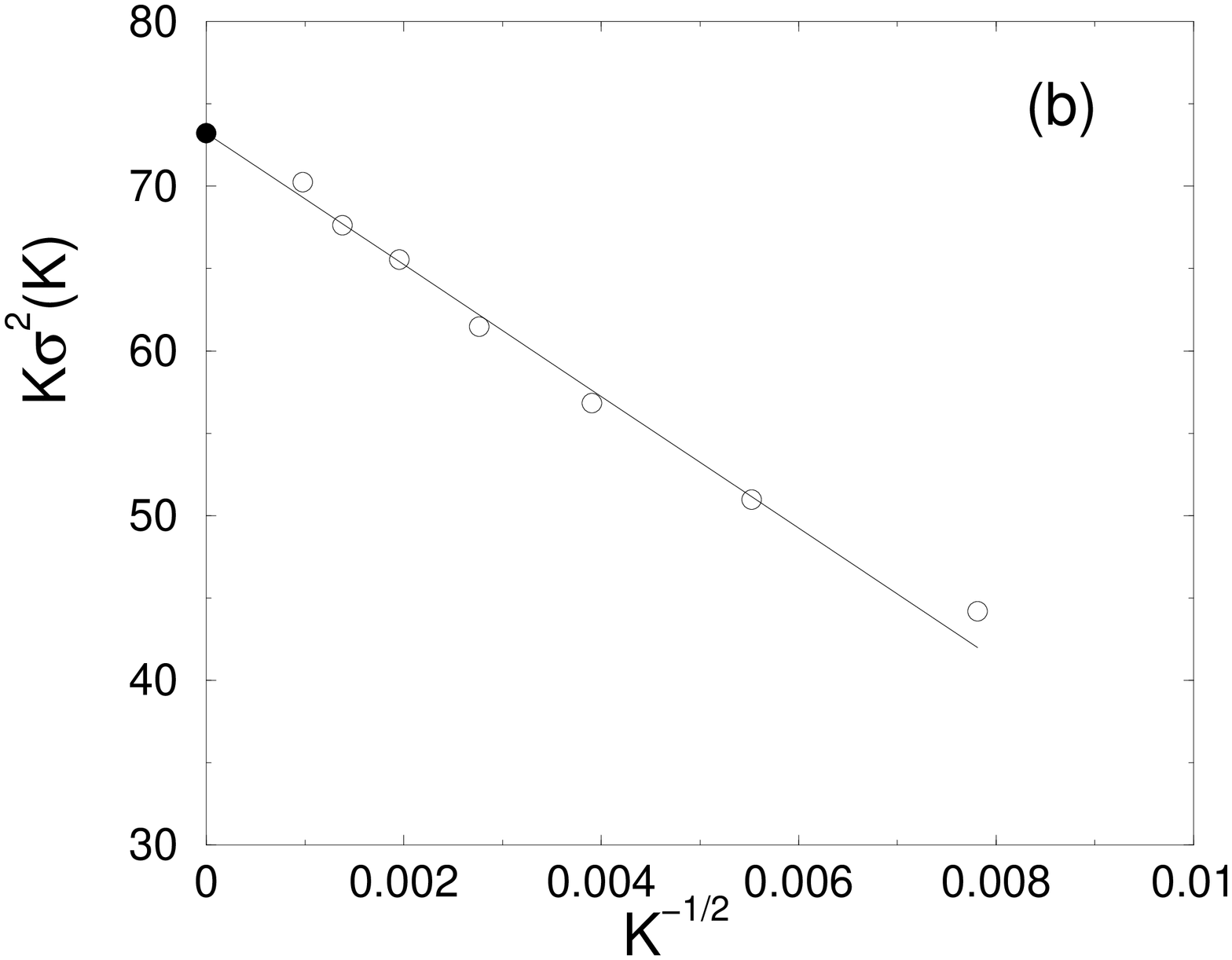}
\caption{Convergence of the parameters of the simulated $P(g|K)$ to the values, 
which follow from the central limit theorem: (a) the mean 
$m(k)$ and (b) the normalized variance $K\sigma^2(K)$.  
In both cases the speed of convergence is $1/\sqrt{K}$ as can be
seen from the straight line fits versus  $1/\sqrt{K}$ with the intercepts equal to the
analytical values $m=0.196$ and $V=73.24$, respectively. The parameters of the simulations
$V_\xi=5.13$ $m_\xi=3.44$, $V_\eta=0.36$, and $m_\eta=0.016$ are taken from the empirical
analysis of the pharmaceutical data base\cite{Fu_PNAS}.} 
\label{conv}
\end{figure}

\begin{figure}
\centering
\includegraphics[scale=0.25,angle=-90]{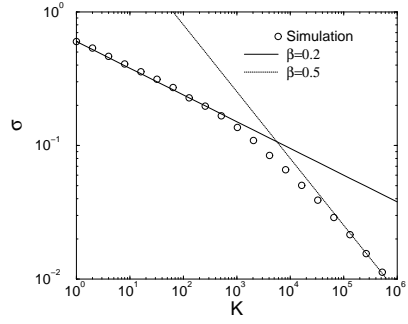}
\caption{Crossover of the size-variance relationship $\sigma(K)$ from $K^{0.2}$ for small
$K$ to $K^{0.5}$ for large $K$. The parameters of the simulations are the same as in Fig.\ref{conv}.}
\label{sigma}
\end{figure}

\begin{figure}
\centering
\includegraphics[scale=0.25,angle=-90]{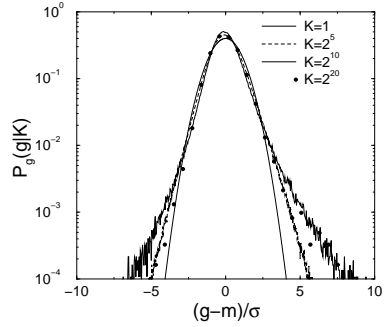}
\caption{Convergence of the shape of the distribution of $P(g|K)$ found in simulations
to limiting Gaussian. One can see the developments of the tent-shape wings as $K$ grows. 
The parameters of the simulations are the same as in Fig.\ref{conv}.}
\label{pg}
\end{figure}

Note, that the convergence of the sum of lognormals to the Gaussian given by Eq. (\ref{nu})
is a very slow process, achieving reasonable accuracy only for $K \gg \mu_\xi(2)\sim \exp(2 V_\xi)$.
For a pharmaceutical database \cite{Fu_PNAS}, we have $V_\xi=5.13$, $m_\xi=3.44$, $V_\eta=0.36$, and 
$m_\eta=0.16$. Accordingly, we can expect convergence only when $K\gg 3\cdot 10^4$.
Figure\ref{conv} demonstrates the convergence of the normalized variance $K\sigma^2(K)$ and mean $m(K)$ of $g$ 
to the theoretical limits given by Eqs. (\ref{e.m}) and (\ref{e.V}) respectively: $V=73.24$ and $m=0.196$. 
In both cases,
the discrepancy between the limiting values and the actual values decreases as $1/\sqrt{K}$.
Interestingly, Eq. (\ref{P_g_large_K}) predicts $\sigma(K) \sim K^{-\beta}$, where $\beta=1/2$.
This value is much larger than the empirical value $\beta\approx 0.2$ observed for the size-variance 
relationships of various socio-economic entities~\cite{Stanley96,Amaral97,Sergey_II,Matia05}. 
However, the slow convergence of $V(K)K$ suggests
that for quite a wide range of $K<1000$, $\sigma(K)\sim K^{-0.2}$ and only at $K>10^4$ there is
a crossover to the theoretical value $\beta=0.5$, (Fig.~\ref{sigma}). Finally, the simulated distribution of $P(g|K)$ has tent-shape wings which develop as $K$
increases (Fig.~\ref{pg}). This feature of the model growth rates may explain the abundance of the 
tent-shaped wings of the growth rates of various systems in nature.  
The most drastic discrepancy between the Gaussian shape and the simulated distribution
$P(g|K)$ can be seen when $K\approx 1000$ and than it starts to decrease slowly, and remains
visible even for $K=10^6$.        

\begin{figure}
\centering
\includegraphics[scale=0.25,angle=-90]{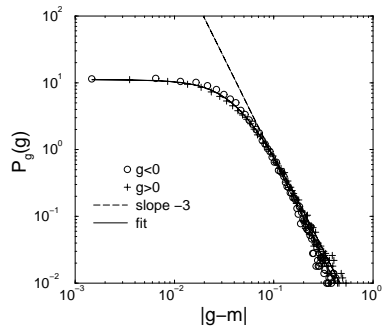}
\caption{Comparison of the shape of the distribution of $P(g)$ for the exponential
distribution of $P(K)=1/\langle K \rangle \exp(-K/\langle K\rangle)$ with $\langle K \rangle =2^{15}$
with the prediction of Eq.(\ref{e.*}). The parameters of the simulation are the same as in
Fig. \ref{conv}. The fitting parameter $V=33$ in Eq. (\ref{e.*}) gives the best agreement
with the simulation results.  One can see a very good convergence to the inverse cubic law
for the wings.}
\label{exp-15}
\end{figure}

Nevertheless, in order to obtain close form approximations for the growth rate,
we will use the Gaussian approximation (\ref{P_g_large_K}) for $P(g|K)$.
The distribution of the growth rate of the old classes can be found by
Eq.~(\ref{P_g_sum}). In order to find a close form approximation, we replace
the summation in Eq.~(\ref{P_g_sum}) by integration and replace the
distributions $P(K)$ by Eq.~(\ref{p_old_class}) and $P(g|K)$ by the
Eq.~(\ref{P_g_large_K}). Assuming $m=0$, we have
\begin{eqnarray}
P_{old}(g) & \approx & {\frac{1 }{\sqrt{2\pi V}}}\int_0^\infty{\frac{1}{K(t)}}\, \exp(
\frac{-K}{K(t)}) \exp(-\frac{g^2\,K}{2\,V})\sqrt{K}\,\,dK, \notag \\ &=&
\frac{\sqrt{K(t)}}{2\,\sqrt{2\,V}} \,\left(
1+\frac{K(t)}{2V}\,g^2\right)^{-\frac{ 3}{2}}, \label{e.*}
\end{eqnarray}
where $K(t)$ is the average number of units in the old classes (see
Eq.~(\ref{mean_K_old})). This distribution decays as $1/g^3$ and thus does
not have a finite variance.  
In spite of drastic assumptions that we make, Eq. (\ref{e.*}) correctly predicts the
shape of the convolution $P_{old}(g)$. Figure~\ref{exp-15} shows the comparison of the 
simulation of the growth rates in the system with the exponential distribution of 
units $P(K)$ with $K(t) =2^{15}$ and the same empirical parameters of the unit size and growth distributions as before. 
The parameter of the analytical 
distribution characterizing its width (variance does not exist), must be taken
$V=33$ which is much smaller than the analytical prediction $V=73.23$. This is not surprising, since for $K=2^{15}$
$K\sigma^2(K)=50$ (see Fig.~\ref{conv}b). Moreover, since we are dealing with the average $\sigma^2(K)K$ for $K<2^{15}$,
we can expect $V<50$. Nevertheless the nature of the power-law wings decaying as $1/g^3$
is reproduced very well.     

For the new classes, when $t\to\infty$ the distribution of number of units
is approximated by
\begin{equation}
P_{\mbox{\scriptsize new}}(K) \approx {\frac{1}{1-b}}K^{-1-\frac{1}{1-b}
}\,\int_0^K\,y^{\frac{1}{1-b}}\,e^{-y}\,\,dy.  \label{e.***}
\end{equation}

Again replacing summation in Eq.~(\ref{P_g_sum}) in the text by integration
and $P(g|K)$ by Eq.~(\ref{P_g_large_K}) and after the switching the order
of integration we have:
\begin{equation}
P_{new}(g) \approx {\frac{1}{1-b}}\,{\frac{1}{\sqrt{2\pi
V}}}\int_0^\infty\,\exp(-y) \,y^{\frac{1}{1-b}}\,dy\,\int_y^\infty\,
\exp(-g^2\,K/2V)\,K^{(-\frac{1}{2}-\frac{1}{1-b}) }\,dK.  \label{e.****}
\end{equation}

As $g\to\infty$, we can evaluate the second integral in Eq.~(\ref{e.****}) by
partial integration:
\begin{eqnarray}
P_{new}(g) & \approx & \frac{1}{1-b}\int_0^\infty \frac{1}{\sqrt{2\pi V}} \,\, \frac{
2V}{g^2} \,\, y^{-\frac{1}{1-b} - \frac{1}{2}} \,\, y^{\frac{1}{1-b}} \,\,
\exp(-y)\,\,\exp(-y\,g^2/2V)\,\,dy, \notag \\ &=& \frac{1}{1-b}\,
\frac{1}{\sqrt{2\pi V}}\,\frac{2V}{g^2} \,\,\frac{1}{
\sqrt{g^2/2V+1}}\,\, \sqrt{\pi} \sim \frac{1}{g^3}.
\end{eqnarray}

We can compute the first derivative of the distribution (\ref{e.****}) by
differentiating the integrand in the second integral with respect to $g$.
The second integral converges as $y\to 0$, and we find the behavior of the
derivative for $g\to 0$ by the substitution $x=K g^2/(2V)$. As $g\to
0$, the derivative behaves as $g\, g^{2[-(3/2)+1/(1-b)]}\sim
g^{2b/(1-b)}$, which means that the function itself behaves as
$C_2-C_1|g|^{2b/(1-b)+1}$, where $C_2$ and $C_1$ are positive constants. For
small $b$ this behavior is similar to the behavior of a Laplace distribution
with variance $V$: $\exp(-\sqrt{2}|g|/\sqrt{V})/\sqrt{2V}=1/\sqrt{2V}-|g|/V$.

When $b\to 0$, Eq.~(\ref{e.****}) can be expressed in elementary
functions:  
\begin{eqnarray}
P_{new}(g)|_{b\to 0} & \approx & \frac{1}{\sqrt{2\pi V}} \, \int_0^\infty
K^{-3/2}\,\exp(-K\,g^2/2\,V)\,dK\,\int_0^K\,\exp(-y)\,y\,\,dy,  \notag \\
& \approx & \frac{1}{\sqrt{2\,V}}\,\,\left(-\frac{1}{\sqrt{1+ g^2/2\,V}} +
\frac{2}{|g|/\sqrt{2\,V} + \sqrt{g^2/2\,V+1 }} \right).
\notag
\end{eqnarray}
Simplifying we find the main result:
\begin{equation}  \label{p_new}
P_{\mbox{\scriptsize new}}(g)|_{b\to 0} \approx \frac{2V}{\sqrt{g^2+2V}\,(|g|+\sqrt{g^2+2V})^2}.
\end{equation}
which behaves for $g\to 0$ as $1/\sqrt{2V}-|g|/V$ and for $g\to\infty$ as
$V/(2g^3)$. Thus the distribution is well approximated by a Laplace
distribution in the body with power-law tails. Because of the discrete nature
of the distribution of the number of units, when $g\gg\sqrt{2V}$ the behavior
for $g\to\infty$ is dominated by $\exp(-g^2/2V)$.

\begin{figure}
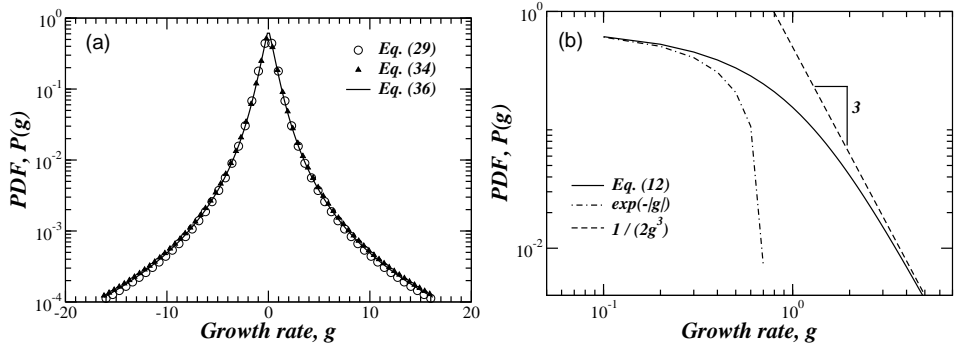

\centering
\includegraphics[scale=0.25]{BuldyrevFig6a.eps}
\includegraphics[scale=0.25]{BuldyrevFig6b.eps}
\caption{ (a) Comparison of three different approximations for the growth
rate PDF, $P(g)$, given by Eq.~(\protect\ref{e.*}), mean field
approximation Eq.~(\protect\ref{e.****}) for $b=0.1$ and
Eq.~(\protect\ref{p_new}). Each $P(g)$ shows similar tent shape behavior
in the central part. We see there is little difference between the three
cases, $b=0$ (no entry), $b=0.1$ (with entry) and the mean field
approximation. This means that entry of new classes ($b>0$) does not
perceptibly change the shape of $P(g)$. Note that we use $K(t)/V_g = 2.16$
for Eq.~(\protect\ref{e.*}) and $V_g=1$ for
Eq.~(\protect\ref{p_new}). (b) The crossover of $P(g)$ given by
Eq.~(\protect\ref{p_new}) between the Laplace distribution in the center
and power law in the tails. For small $g$, $P(g)$ follows a Laplace
distribution $P(g) \sim \exp(-|g|)$, and for large $g$, $P(g)$
asymptotically follows an inverse cubic power law $P(g) \sim g^{-3}$.}
\label{crossover}
\end{figure}

In Fig.~\ref{crossover}a we compare the distributions given by
Eq.~(\protect\ref{e.*}), the mean field approximation
Eq.~(\protect\ref{e.****}) for $b=0.1$ and Eq.~(\protect\ref{p_new}) for
$b\to 0$. We find that all three distributions have very similar tent shape
behavior in the central part. In Fig.~\ref{crossover}b we also compare the
distribution Eq.~(\protect\ref{p_new}) with its asymptotic behaviors for
$g\to 0$ (Laplace cusp) and $g\to \infty$ (power law), and find the crossover
region between these two regimes.

\section{ Conclusions }

The analytical solution of this model can be obtained only for certain
limiting cases but a numerical solution can be easily computed for any set of
assumptions. We investigate the model numerically and analytically (see
and find:

\begin{itemize}

\item[{(1)}] In the presence of the influx of new classes ($b>0$), the
distribution of units converges for $t\to \infty$ to a
power law $P(K)\sim K^{-\varphi}$, $\varphi=2+b/(1-b)\ge 2$.  Note
that this behavior of the power-law probability density function leads
to the power law rank-order distribution where rank of a class $R$ is
related to the number of its units $K$ as
\begin{equation}
R=N(t)\int_K^\infty P(K)dk\sim K^{-\varphi+1}.
\end{equation} 
Thus $K\sim R^{-\zeta}$,
where $\zeta=1/(\varphi-1) = 1-b \leq 1$, which leads in the limit $b\to 0$ to the celebrated 
Zipf's law\cite{Zipf49} for cities populations, $K\sim 1/R$. Note that this equation
can be derived for our model using elementary considerations. Indeed, due to proportional 
growth the rank of a class, $R$, is proportional to the time of its creation $t_0$. The number
of units $n(t_0)$ existing at time $t_0$ is also proportional to $t_0$ and thus also proportional
to $R$. According to the proportional growth, the ratio of the number of 
units in this class to the number of units
in the classes existed at time $t_0$ is constant: $K(t_0,t)/n_e(t_0,t) = 1/n(t_0)$. 
If we assume that the amount of units in the classes, created after $t_0$ can be neglected 
since the influx of new classes $b$ is small, we can approximate $n_e(t_0,t) \approx n(t)\sim t$.
Thus for large $t$, $n_e(t_0,t)$ is independent of $t_0$ and hence $K(t_0,t) \sim 1/R$. 
If we do not neglect the influx of new classes, Eq. (\ref{n_e}) gives 
$n_e(t_0,t)\sim t_0^b$, hence $K(t_0,t)\sim 1/R^{1-b}$.

\item[{(2)}] The conditional distribution of the logarithmic growth
  rates $P(g|K)$ for the classes consisting of a fixed number $K$ of
  units converges to a Gaussian distribution (\ref{P_g_large_K}) for
  $K\to\infty$. Thus the width of this distribution, $\sigma(K)$, decreases as
  $1/K^\beta$, with $\beta=1/2$. Note that due to slow convergence of
  lognormals to the Gaussian in case of wide lognormal distribution of unit
  sizes $V_\xi=5.13$, computed from the empirical data~\cite{Fu_PNAS}, we have
  $\beta=0.2$ for relatively small classes.  This result is consistent
  with the observation that large firms with many production units
  fluctuate less than small firms~\cite{Sutton97,Amaral97,Amaral98,Hymer}. 
  Interestingly, in case of large $V\xi$, $P(g|K)$ converges
  to the Gaussian in the central interval which grows with $K$,  but outside this
  interval it develops tent-shape wings, which are becoming increasingly wider, 
  as $K\to \infty$. However, they remain limited by the distibution of the logarithmic
  growth rates of the units, $P_\eta (\ln \eta)$.    

\item[{(3)}] For $g\gg V_{\eta}$, the distribution
  $P(g)$ coincides with the distribution of the logarithms of the growth
  rates of the units:
  \begin{equation}
    P(g)\approx P_\eta(\ln\eta).
  \end{equation}
  In the case of power law distribution $P(K)\sim K^{-\varphi}$ which dramatically
  increases for $K\to 1$, the distribution $P(g)$ is dominated by the growth
  rates of classes consisting of a single unit $K=1$, thus the distribution
  $P(g)$ practically coincides with $P_{\eta}(\ln\eta)$ for all $g$. Indeed,
  empirical observations of Ref.~\cite{Fu_PNAS} confirm this result.

\item[{(4)}] If the distribution $P(K)\sim K^{-\varphi}$, $\varphi>2$ for
  $K\to\infty$, as happens in the presence of the influx of new units $b\neq
  0$, $P(g)=C_1-C_2|g|^{2\varphi-3}$, for $g\to 0$ which in the limiting case
  $b\to 0$, $\varphi\to 2$ gives the cusp $P(g)\sim C_1-C_2|g|$ ($C_1$ and
  $C_2$ are positive constants), similar to the behavior of the Laplace
  distribution $P_{\rm L}(g)\sim \exp(-|g|C_2)$ for $g\to 0$.

\item[{(5)}] If the distribution $P(K)$ weakly depends on $K$ for $K\to 1$,
  the distribution of $P(g)$ can be approximated by a power law of $g$:
  $P(g)\sim g^{-3}$ in wide range $\sqrt{V_g/K(t)}\ll g\ll\sqrt{V}$, where
  $K(t)$ is the average number of units in a class. This case is realized for
  $b=0$, $t\to\infty$ when the distribution of $P(K)$ is dominated by the
  exponential distribution and $K(t)\to\infty$ as defined by
  Eq.~(\protect\ref{P_K_old}). In this particular case, $P(g)$ for $g\ll
  \sqrt{V_g}$ can be approximated by Eq.(\ref{e.*})

\item[{(6)}] In the case in which the distribution $P(K)$ is not dominated by
one-unit classes but for $K\to\infty$ behaves as a power law, which is the
result of the mean field solution for our model when $t\to\infty$, the
resulting distribution $P(g)$ has three regimes, $P(g)\sim
C_1-C_2|g|^{2\varphi-3}$ for small $g$, $P(g)\sim g^{-3}$ for intermediate $g$,
and $P(g)\sim P(\ln\eta)$ for $g\to\infty$. The approximate solution of
$P(g)$ in this case is given by Eq.~(\ref{e.****}) 
For $b\neq 0 $ Eq.~(\ref{e.****}) can not be expressed in elementary
functions. In the $b\to 0$ case, Eq.~(\ref{e.****}) yields the main
result Eq.(\ref{p_new}).
which combines the Laplace cusp for $g\to 0$ and the power law decay for
$g\to\infty$. Note that due to replacement of summation by integration in
Eq.~(\ref{P_g_sum}), the approximation Eq.~(\ref{p_new}) holds only for
$g<\sqrt{V_{\eta}}$.
\end{itemize}  
In conclusion we want to emphasize that although the derivations of the
distributions (\ref{e.*}), (\ref{e.****}), and (\ref{p_new}) are not rigorous they satisfactory reproduce the
shape of empirical data, especially the $1/g^3$ behavior of the wings of the
distribution of the growth rates and the sharp cusp near the center.



\newpage 

\end{document}